\documentclass[manuscript]{aastex62}

\newcommand{\gaia}{\texttt{Gaia}}
\usepackage{CJK}
\usepackage{amsmath}
\usepackage{subfigure}
\graphicspath{{./}{figures/}}

\submitjournal{ApJS}
\received{October 28, 2022}
\revised{February 6, 2023}
\accepted{February 23, 2023}
\shorttitle{M-giant stars catalog}
\shortauthors{Li et al.}

\begin{document}
\begin{CJK*}{UTF8}{gbsn}
\title{ Value-added catalog of M-giant stars in LAMOST DR9}

\correspondingauthor{Jing Li}
\email{lijing@shao.ac.cn; jzhong@shao.ac.cn}

\author[0000-0002-4953-1545]{Jing Li （李静）}
\affiliation{School of Physics and Astronomy, China West Normal University,
1 ShiDa Road, Nanchong 637002, China}

\author[0000-0003-3787-0790]{lin Long （龙琳）}
\affiliation{School of Physics and Astronomy, China West Normal University,
1 ShiDa Road, Nanchong 637002, China}

\author[0000-0001-5245-0335]{Jing Zhong (钟靖)}
\affiliation{Key Laboratory for Research in Galaxies and Cosmology, Shanghai Astronomical Observatory, Chinese Academy of Sciences, 80 Nandan Road, Shanghai 200030, China}

\author[0000-0001-6395-2808]{lin Tang （唐林）}
\affiliation{School of Physics and Astronomy, China West Normal University,
1 ShiDa Road, Nanchong 637002, China}
\author[0000-0002-6434-7201]{Bo Zhang （章博） }
\affiliation{Key Laboratory of Space Astronomy and Technology, National Astronomical Observatories, Chinese Academy of Sciences, 
Beijing 100101, China}

\author[0000-0003-3713-2640]{Songmei Qin (秦松梅)}
\affiliation{Key Laboratory for Research in Galaxies and Cosmology, Shanghai Astronomical Observatory, Chinese Academy of Sciences, 80 Nandan Road, Shanghai 200030, China}
\affiliation{School of Astronomy and Space Science, University of Chinese Academy of Sciences, No. 19A, Yuquan Road, Beijing 100049, China}

\author{Yirong Chen （陈易容）}
\affiliation{School of Physics and Astronomy, China West Normal University,
1 ShiDa Road, Nanchong 637002, China}

\author{Zhengzhou Yan （闫正洲）}
\affiliation{School of Physics and Astronomy, China West Normal University,
1 ShiDa Road, Nanchong 637002, China}

\author{Li Chen (陈力)}
\affiliation{Key Laboratory for Research in Galaxies and Cosmology, Shanghai Astronomical Observatory, Chinese Academy of Sciences, 80 Nandan Road, Shanghai 200030, China}
\affiliation{School of Astronomy and Space Science, University of Chinese Academy of Sciences, No. 19A, Yuquan Road, Beijing 100049, China}

\author[0000-0002-0642-5689]{Xiang-Xiang Xue(薛香香)}
\affiliation{Key Laboratory of Optical Astronomy,
National Astronomical Observatories, Chinese Academy of  Sciences, Beijing 100101, China}
\affiliation{School of Astronomy and Space Science, University of Chinese Academy of Sciences, 19A Yuquan Road, Shijingshan District, Beijing 100049, China}
\author{Jinliang Hou (侯金良)}
\affiliation{Key Laboratory for Research in Galaxies and Cosmology, Shanghai Astronomical Observatory, Chinese Academy of Sciences, 80 Nandan Road, Shanghai 200030, China}
\affiliation{School of Astronomy and Space Science, University of Chinese Academy of Sciences, No. 19A, Yuquan Road, Beijing 100049, China}

\author[0000-0002-0349-7839]{Jian-Rong Shi （施建荣） }
\affiliation{Key Laboratory of Optical Astronomy,
National Astronomical Observatories, Chinese Academy of  Sciences, Beijing 100101, China}
\affiliation{School of Astronomy and Space Science, University of Chinese Academy of Sciences, No. 19A, Yuquan Road, Beijing 100049, China}



\begin{abstract} 
In this work, we update the catalog of M-giant stars from the low-resolution spectra of the Large Sky Area Multi-Object Fiber Spectroscopic Telescope (LAMOST) Data Release 9. There are 58,076 M giants identified from the classification pipeline with seven temperature subtypes from M0 to M6. The 2471 misclassified non-M-giant stars are white dwarf binaries, early types, and M dwarfs. And the contamination rate is $4.2\%$ in the M-giants sample. A total of 372 carbon stars were identified by CaH spectral indices, and were further confirmed by the LAMOST spectra.  We update the correlation between the $(W1-W2)_0$ color and [M/H] from APOGEE DR17. We calculate the radial velocities of all M giants by applying cross-correlation to the spectra between 8000 and 8950 \AA~with synthetic spectra from ATLAS9. Taking star distances less than 4 kpc from Gaia EDR3 as the standard, we refitted the photometric distance relation of M giants.  And based on our M-giant stars, we select a group of Sagittarius stream members, whose sky and 3D velocity distributions are well consistent with K-giant Saggitarius stream members found in Yang et al. With our M giants, we find that the disk is asymmetric out to R = 25 kpc, which is 5 kpc further out than detected using K giants. 
\end{abstract}


\section{Introduction}
\label{sec:intro}
M giants are red giant branch (RGB) stars with low temperature ($<$4000K) and high luminosity \citep[log $L/L_{\sun}$ $\sim$ 3-4;][]{2009ssc..book.....G}, which enables them are suitable as tracers for discovering and identifying the substructures in the Galactic outer disk and remnants of stellar streams in the Galactic halo. Over the last 20 yr, a great number of large-scale survey projects, e.g., the Two Micron All Sky Survey (2MASS), UKIRT Infrared Deep Sky Survey (UKIDSS), Wide-field Infrared Survey Explorer (WISE), Pan-STARRS, and \gaia{}, gathered thousands of photometric M-giant stars, which provided key support for the study of substructures, especially for the Sagittarius stream. For example, M giants were first used to map out the global view of the Sagittarius Dwarf galaxy \citep{2003ApJ...599.1082M} with 2MASS M giants. Using UKIDSS, \citet{2014AJ....147...76B} established a distant M-giant sample to explore the history of Sagittarius accretion in the outer halo of our galaxy.

Compared to the photometric M-giants data, spectral data can provide radial velocity and chemical abundances parameters, which helps to study the structure and evolution of the Milky Way. The Large Sky Area Multi-Object Fiber Spectroscopic Telescope \citep[LAMOST;][]{1996ApOpt..35.5155W,Cui2012,2004ChJAA...4....1S,Luo2012,Zhao2012,2022Innov...300224Y} provided the possibility to select a large number of relative pure M-giant spectra.

As the most efficient spectroscopic survey telescope, LAMOST, also named the Guo Shou Jing Telescope, is a 4 m quasi-meridian reflective Schmidt telescope with 4000 fibers \citep{Cui2012,Luo2012,Zhao2012}. From 2011 to 2018, it has achieved its first stage low spectral resolution ($R \sim 1800$) survey and obtained more than 10 million spectra, including about 9 million stellar spectra \citep{Cui2012,2012RAA....12..735D,Luo2012,Zhao2012}.

Since 2018 October, LAMOST has started the second stage survey program, LAMOST II, which contains both low- and medium-resolution spectroscopic surveys. LAMOST II is taking around $50\%$ of all nights to continue the low-resolution spectroscopic survey. The other $50\%$ of nights are assigned to a medium-resolution ($R \sim 7500$) survey \citep{2020arXiv200507210L}. As of 2022 March, LAMOST DR9 published more than 11 million low-resolution spectra, covering the wavelength range from 3690 to 9100 \AA. Meanwhile LAMOST II released nearly 30 million medium-resolution spectra covering the wavelength range from 4950 to 5350 \AA~through the blue camera and 6300 to 6800 \AA~through the red camera \citep{2020arXiv200507210L}. 

In this work, we present a much larger catalog of spectroscopic M giants with LAMOST Data Release 9 (DR9) low-resolution spectra. The paper is organized as follows. In section 2, we describe how we select M giants from LAMOST DR9. In section 3, we describe the separation between M giants, M dwarfs, and carbon stars using photometric and spectroscopic index features. In section 4, we analyze the properties of these M giants, reconstructing the photometric relations for metallicity and distance, and calculating radial velocities. In section 5, we detect the properties of the Saggitarius stream (Sgr) and the Galactic outer disk with our M giants. A summary is presented in the last section.

\section{Candidates Selection in LAMOST DR9}
\subsection{LAMOST M Giants Selection History }
In order to recognize M-type stars in the early LAMOST database, \citet{2015AJ....150...42Z} developed an automatic template-fitting algorithm to classify M-type stars, which normalized the LAMOST pseudo-continuum in the range of 6000$-$8000 \AA~and performed $\chi^{2}$ fitting between object spectra and M-dwarf template spectra. The template spectrum that corresponds to the minimum $\chi^{2}$ value was defined as the best-fit template.

Then, \citet{2015RAA....15.1154Z} found that M giants and M dwarfs can be well discriminated in the CaH2+CaH3 versus TiO5 spectral indices diagram and assembled a set of M-giant templates with spectral subtypes ranging from M0 to M6. Finally, a total of 8,639 M giants and 101,690 M dwarfs were positively classified by the template-fitting algorithm from LAMOST DR1. 

\citet{2016RAA....16..125L} applied the same method to LAMOST DR2 and expanded the M-giants sample to 21,696. Furthermore, \citet[hereafter Li16]{2016ApJ...823...59L} found that the combination of WISE+2MASS \citep{2006AJ....131.1163S,2010AJ....140.1868W} bands was more efficient in selecting M-giant samples than the 2MASS bands alone. Meanwhile, Li16 estimated the photometric metallicity of M giants and further derived the photometric distances calibrated by Sgr stream, LMC, and SMC members.   

\citet[hereafter Z19]{2019ApJS..244....8Z} applied the template-fitting algorithm to LAMOST DR5, and found 39,796 M giants. In addition, \citet{2022ApJS..260...45D} adopted the ULySS package \citep{2009A&A...501.1269K} to perform $\chi^{2}$ minimization with model spectra generated from the MILES interpolator \citep{2006MNRAS.371..703S,2011A&A...532A..95F} and determined the stellar atmospheric parameters for the M-type stars from LAMOST DR8. And a catalog of 763,136 spectra corresponding to 616,314 M-type stars with derived stellar parameters was presented by them.

\subsection{M Giants from LAMOST DR9}
In this work we have adopted the same template-fitting pipeline combined with a classification algorithm \citep{2015AJ....150...42Z} and revised M-type spectral templates \citep{2015RAA....15.1154Z} to identify and classify M-type stars from LAMOST DR9 low-resolution spectra. The M-giant templates span spectral subtypes from M0 to M6. 
Our method can efficiently select M giants from LAMOST low-resolution spectra. But the spectral subtypes of our templates only extend to M6 because the LAMOST spectra do not containing a large-enough metal-poor sample to make templates for the other subtypes.

\begin{figure*}
  \centering
   \includegraphics[angle=0,scale=0.6]{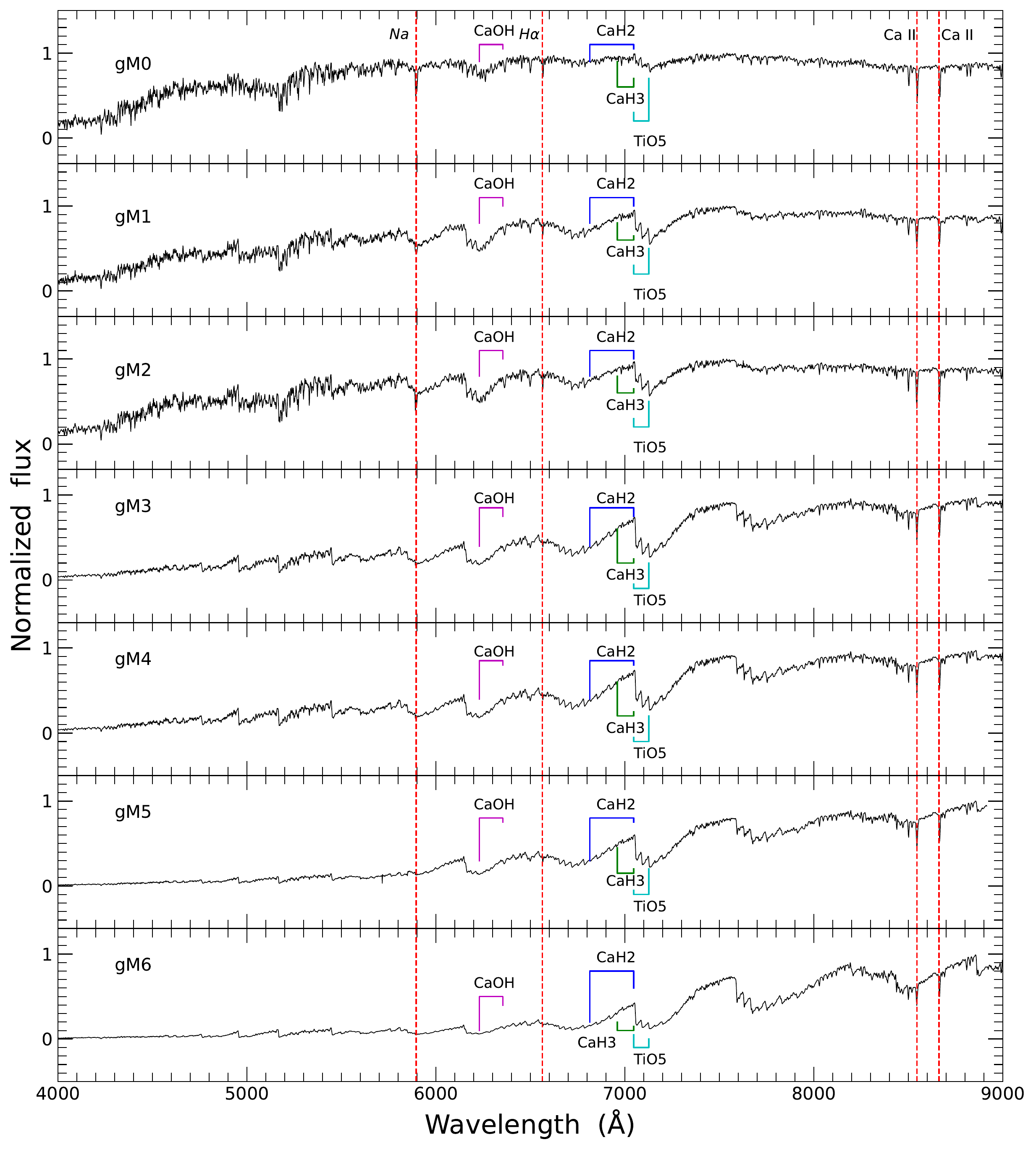}
  \caption{ The M-giant templates from M0 to M6.}
   \label{mgspec}
\end{figure*}

In Figure~\ref{mgspec}, we present the M-giant template spectra, which have marked molecular and atomic spectral features, e.g., CaH2 (6814-6846 \AA), CaH3 (6960-6990 \AA), TiO5 (7126-7135  \AA), CaOH (6230-6240 \AA), \ion{Na}{1} (8172-8197 \AA), \ion{Ca}{2} (8484-8662 \AA), and H$_\alpha$ \citep[6563 \AA;][]{1995AJ....110.1838R,2015RAA....15.1154Z}. From top to bottom, the spectra are presented along their temperature sequence. As before, although the optimized template-fitting algorithm largely eliminates the majority of noisy spectra in LAMOST, a small fraction of weird spectra still remain in the M-type star sample; most of them suffer low signal-to-noise ratios (S/Ns) or serious sky line contamination. To purify the M-type star sample further, candidates have to meet the following two additional criteria (Z19): (i) the mean S/N in the range 6000$-$8000 \AA~must be greater than 5 and (ii) the spectral indices must be located on the M-type star locus (0 $<$ TiO5 $<$ 1.2 and 0.6 $<$ CaH2+CaH3 $<$ 2.4). After excluding outlier spectra and combining duplicated spectra, a total of 822,752 spectra were identified as M-type stars, including 58,076 M-giant spectra and 764,676 M-dwarf spectra. In this work, we select M-giant stars from the LAMOST DR9 low-resolution spectra. The M giants selected from median-resolution spectra and the M-dwarf stars will be discussed in our next work.

\section{Data Analysis and Classification Results}
\subsection{M Giants}
\label{results}
\label{mstar}
\begin{figure*}
  \centering
  \includegraphics[angle=0,scale=0.45]{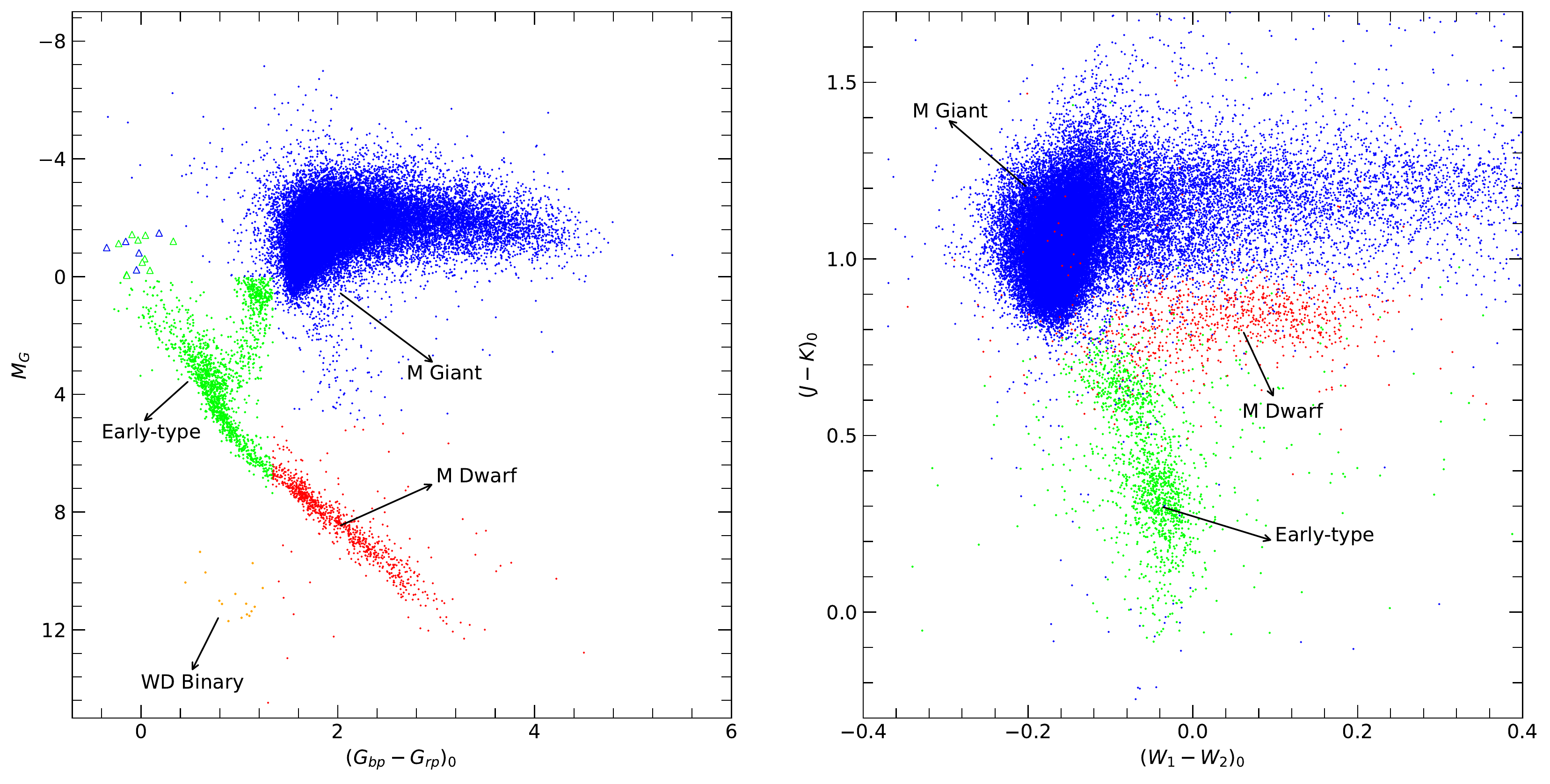}
  \caption{Left: $G$-band absolute magnitude as a function of $(G_{bp}-G_{rp})_0$ color. Right: dereddened $(J-K)_0$ vs. $(W1-W2)_0$ color-color diagram. Note that all dots are M-giant candidates that were classified through LAMOST spectra. We use different  colors to distinguish misidentified stars, which include early-type stars (green), M dwarfs (red) and white dwarf binaries (orange).}
  \label{cmd}
\end{figure*}

First we check the distribution of M-giant and M-dwarf candidates selected by our pipeline with spectral indices (CaH2+CaH3 versus TiO5) and infrared colors as described in \citet{2015RAA....15.1154Z}, Li16 and Z19. To derive the intrinsic $(J-K)_0$ and $(W1-W2)_0$ colors of each M giant, we adopted Galactic reddening $E(B-V)$ from the 3D dust maps of \citet{2019ApJ...887...93G}, in combination with the extinction coefficients determined by \citet{2013MNRAS.430.2188Y}. Then we plot the spectral indices diagram and the dereddened $(J-K)_0$ versus $(W1-W2)_0$ color-color diagram. As expected, the majority of M giants and M dwarfs are well separated in both the spectroscopic and photometric parameter space, which are same as the Figure 1 of Z19.

Then we calculate the absolute magnitude $M_G$ and intrinsic $(G_{bp}-G_{rp})_0$ color of each M-giant star. The distance moduli and 3D dust maps were determined by \citet{2021AJ....161...147B} and \citet{2019ApJ...887...93G}, while the extinction $A_G$ and reddening $E(BP-RP)$ were calculated as $A_G=2.74 \times E(B-V)$ and $E(BP-RP)=1.339 \times E(B-V)$ \citep{2018MNRAS.479L.102C}. Figure~\ref{cmd} shows the distribution of M giants which were identified by our template-fitting algorithm. In the left panel, stellar locations in the $M_G$ versus  $(G_{bp}-G_{rp})_0$ diagram  \citep{2021A&A...650C...3G} indicate that there are indeed contaminants in the M-giants sample, including early-type stars, M dwarfs, and a few white dwarfs, as discussed in Z19 for LAMOST DR5 M giants. Moreover, we check the LAMOST spectra of the stars marked by triangles, confirming the green triangles are early-type stars and the stars marked by blue triangles are intrinsic M giants. We also plot the dereddened $(J-K)_0$ versus $(W1-W2)_0$ color-color diagram of M giants in the right panel. Comparing with the distribution of K/M stars in Li16 and Z19, the location of each stellar component is consistent with the reported region. Through close visual inspection of the LAMOST spectra for each of these non-M giants, we also find that a large fraction of identified early-type star candidates are main-sequence M-giant binaries, and all labeled as white dwarfs are white dwarf-M dwarf binaries. In particular, the early-type stars in the M-giants sample are labeled as non-M-type stars, the M dwarfs are labeled as M dwarfs, and the white dwarfs are labeled as white dwarf binaries in the M-giants catalog (see Section~\ref{catalog}).

\begin{figure*}
  \centering
   \includegraphics[angle=0,scale=0.8]{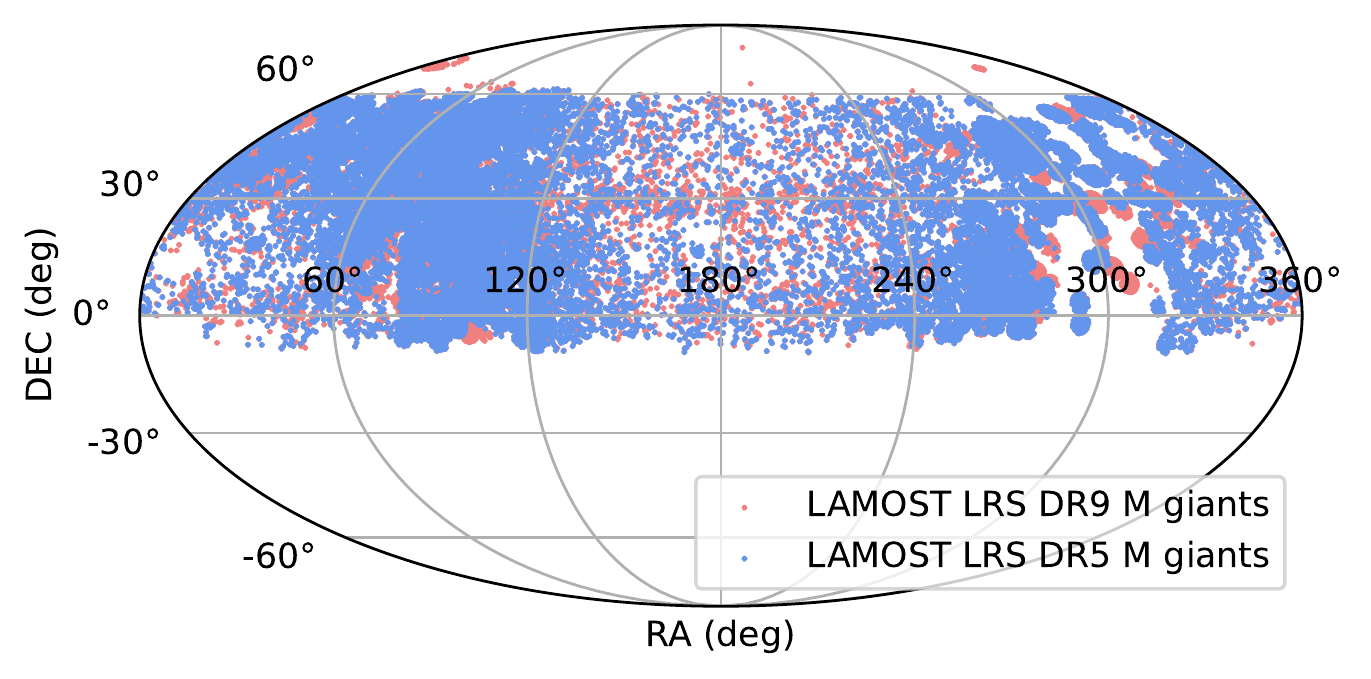}
  \caption{The sky distribution of M-giants sample in equatorial coordinate. The red dots represent our M-giants sample from LAMOST DR9, the blue dots represent the M-giants sample from LAMOST DR5.}
   \label{ra}
\end{figure*}

Finally, the total number of non-M-type stars are 2471 (including white dwarf binaries), corresponding to a contamination rate of about 4.2$\%$ in the M-giants sample, which is similar to the contamination rate (4.6$\%$) reported in Z19. Figure~\ref{ra} shows the sky distribution of our M giants.

\subsection{Carbon stars}
Carbon stars, first recognized by \citet{1869AN.....73..129S}, are peculiar objects with optical spectra characterized by strong carbon molecular bands (CH, CN, or C$_2$). Compared to normal stars, carbon stars show an inversion of the C/O ratio (C/O $>$ 1). \citet{1996ApJS..105..419B} revised the MK classification criteria and classified carbon stars into five types, i.e., C-H, C-R, C-J, C-N, and barium stars. And carbon stars are excellent kinematics tracers of galaxies. They can also be served as visible standard candles for galaxies.

Using the LAMOST spectroscopic data, carbon stars have been systematically investigated by a few authors. \citet{2015RAA....15.1671S} found 183 carbon stars from the LAMOST pilot survey with an  efficient manifold ranking algorithm. In the LAMOST DR2 catalog, \citet{2016ApJS..226....1J} reported 894 carbon stars, using a series of spectral indices (C$_2$ at 5635 \AA, \ion{Ba}{2} at 4554 \AA, and CN at 7065 and 7820 \AA) as selection criteria. Using an efficient machine-learning algorithm, \citet{2018ApJS..234...31L} presented a catalog of 2651 carbon stars from LAMOST DR4. Z19 classified 289 carbon stars from LAMOST DR5 misclassified as M giants, where there are 224 common stars with the published carbon stars catalog. 

There are also some researches about the chemical and kinematic properties of carbon stars. Based on \gaia{} data, \citet{2020CoBAO..67..206K} studied 127 C stars detected on the First Byurakan Spectral Sky Survey (FBS), from which 56 are N$-$type AGB stars and 71 are CH$-$type giants. All FBS-detected C stars are giants and AGB stars in the Galactic halo. \citet{2020A&A...633A.135A} analyzed \gaia{} DR2 data for a sample of 210 luminous red giant carbon stars of N, SC, J, and R hot spectral types belonging to the solar neighborhood and fulfilling the criterion $\epsilon(\pi)/\pi \leq 0.2$, in order to derive accurate luminosities and kinematic properties. And \citet{2022A&A...664A..45A} reported the identification of ∼2660 new carbon stars candidates that they identified through 2MASS photometry, Gaia astrometry, and their location in the \gaia{}$-$2MASS diagram.

\begin{figure*}
  \centering
   \includegraphics[angle=0,scale=0.8]{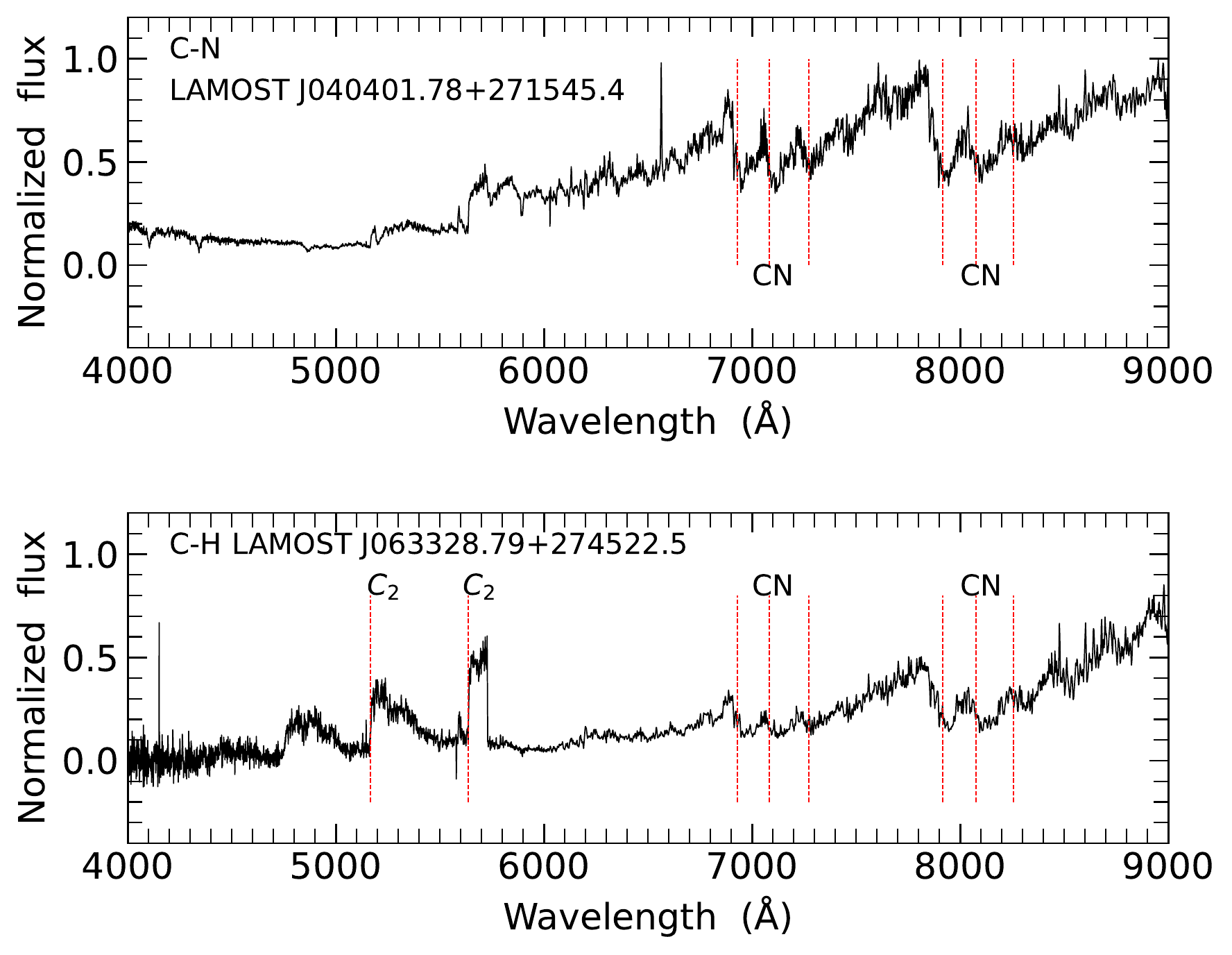}
  \caption{Spectrum examples of C-N and C-H stars in our LAMOST-derived catalog.}
   \label{cstar}
\end{figure*}

Many carbon stars are considered asymptotic giant branch stars undergoing the third dredge-up process (Z19). We believe that a small number of late-type carbon stars could be included in the M-type catalog and are misclassified as M giants in our sample because of their low surface temperatures. And we selected carbon-star candidates whose spectral indices satisfy the criteria [CaH3 $-$ 0.8 $\times $ CaH2 $-$ 0.1] $<$ 0 (Z19). Of the 404 carbon-star candidates that passed the selection criteria, 372 are confirmed as carbon stars by visual inspection, while most of the excluded stars have low-S/N spectra. The fact that we find two carbon-star spectra types, C-N and C-H stars, the same as the work in Z19. And we show the two types of carbon spectra examples in Figure~\ref{cstar}.

\section{Properties of M Giants}
\subsection{Metallicity}

\begin{figure*}
   \centering
   \includegraphics[angle=0,scale=0.8]{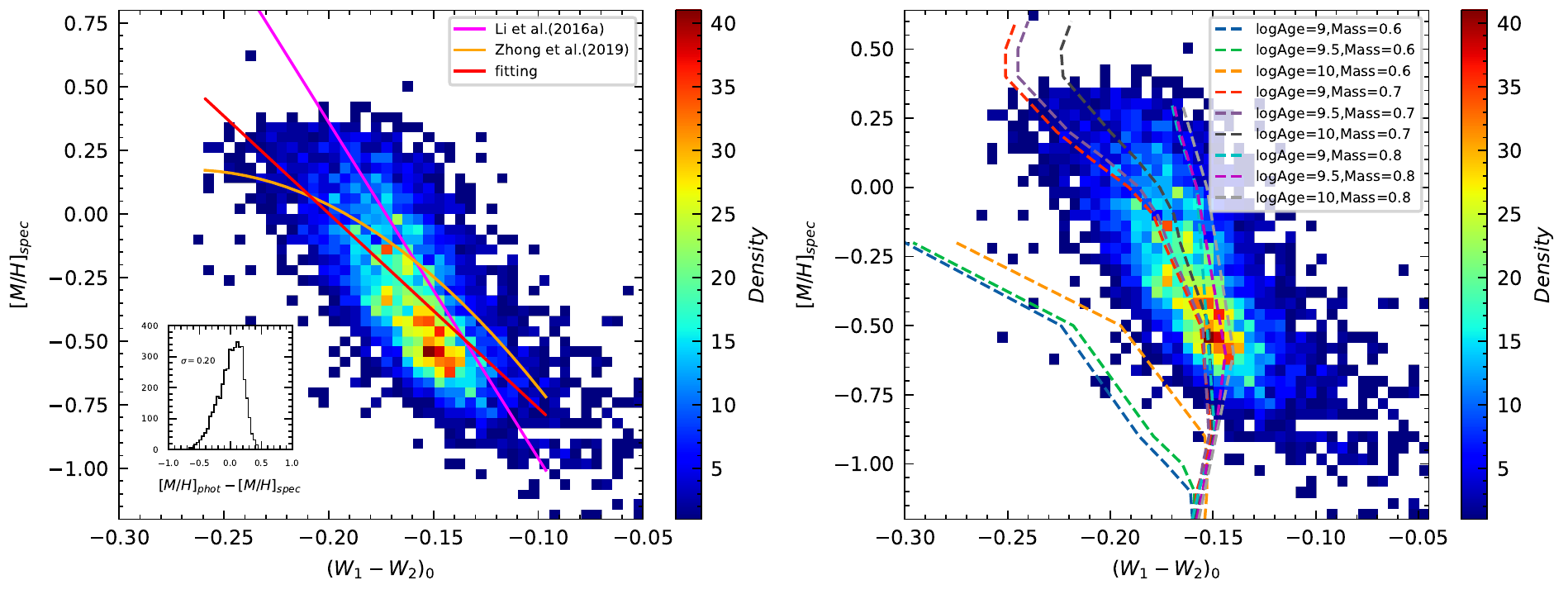}
  \caption{Left: metallicity distribution of APOGEE M giants vs. $(W1-W2)_0$ color. The red line shows the best-fit relationship. Right: the lines correspond to log(age) = 9, 9.5, 10 PARSEC isochrones with mass = 0.6, 0.7, and 0.8 $M_{\odot}$, respectively.}
   \label{mh}
\end{figure*}

The Sloan Digital Sky Survey (SDSS) project APOGEE \citep{holtzman2015} is taking high-S/N and high-resolution (R = 22,500) near-infrared (NIR) spectra, resulting in detailed chemistry and a measurement of [M/H] to a precision of better than 0.1 dex. As in Li16 and Z19, we cross-matched our M-giants sample with APOGEE DR17, finding 4599 stars in common, which provides a much larger sample than used for fitting in Li16 (268 common stars with APOGEE DR12) and Z19 (2689 common stars with APOGEE DR14). We hope to find a higher accuracy color$-$metallicity relation in our work with the APOGEE updated metallicities. 
Figure~\ref{mh} (left panel) shows the APOGEE metallicities of the resulting sample of 4599 M giants as a function of $(W1-W2)_0$ color. The red line shows the fitting curve in this work. The correlation with $(W1-W2)_0$ can be fitted with the following polynomial relation:
\begin{equation}
    [M/H]=-7.647*(W1-W2)_0-1.528.
\end{equation}
As can be seen from the inset figure, the residual scatter about this relation is 0.20 dex. We compare the new result to the previous similar work in Li16 shown as a pink line and Z19 shown as an orange curve. It should be noted that our fitting is only effective in the $(W1-W2)_0$ range from $\sim-0.26$ to $\sim-0.1$ dex. 

In the right panel of Figure~\ref{mh}, we plot the theoretical isochrones from the PAdova and TRieste Stellar Evolution Code \citep[PARSEC;][]{2012MNRAS.427..127B} for our sample. As illustrated in the right panel, a large proportion of stars lie in the coverage of isochrones.

\subsection{Radial Velocities}

The radial velocity is calculated by using the cross-correlation-based $laspec$ algorithm derived by \citet{2021ApJS..256...14Z}. It is applied to the spectra of M-giant stars with wavelengths between 8350 and 8800 \AA~from LAMOST DR9, where the \ion{Ca}{2} triplet is included in this band. The cross-correlation function \citep[CCF;][]{1979AJ.....84.1511T} is evaluated by the following equation:
\begin{equation}
CCF(v|F,G)=\frac{Cov(F,G(v))}{\sqrt{Var(F)Var(G(v))}}.
\end{equation}
where $F$ is the normalized observed spectrum of a given M-giant star and $\boldsymbol{G}$ is the synthetic spectrum from ATLAS9 \citep{2018A&A...618A..25A}. 

\begin{figure*}
  \centering
   \includegraphics[angle=0,scale=0.55]{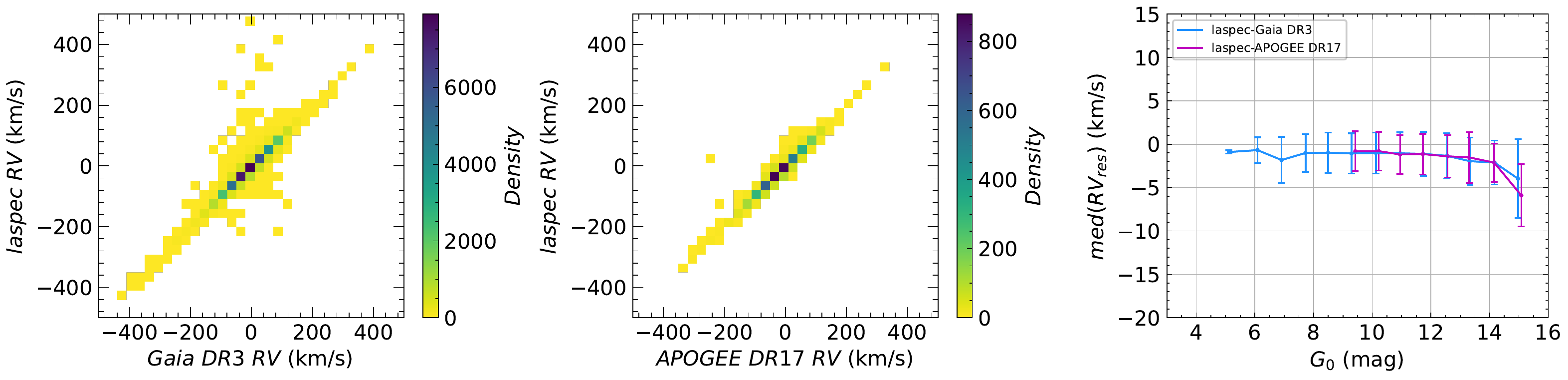}
   \caption{Left and median panel: the density map of \gaia{} DR3 and APOGEE DR17 radial velocities vs. our radial velocities. Right panel: median residuals of the radial velocity as a function of $G_0$ mag. The error bars represent the 68.3$\%$ confidence intervals on the measurements of the medians.}
   \label{rv}
 \end{figure*}

To evaluate our results, we cross-matched our data with \gaia{} DR3 \citep{2022arXiv220800211G,2022arXiv220605902K} and APOGEE DR17 \citep{2022ApJS..259...35A}, finding 43,145 and 4587 common stars, respectively. Stellar spectra for which the \ion{Ca}{2} triplet lines are unapparent and which have abnormal fluxes are excluded. Figure~\ref{rv} shows the comparison. The velocity offset and scatter with Gaia DR3 and APOGEE DR17 are around $\sim$1 km s$^{-1}$ and $\sim$4.6 km s$^{-1}$, respectively. The right panel shows the median residuals of the radial velocities as a function of $G_0$ mag. Our result presents a systematic difference, which reaches approximately $\sim$1 km s$^{-1}$ when $G_0$ is fainter than 14 mag and $\sim$5 km s$^{-1}$ when $G_0$ mag is brighter than 14 mag. 
The Gaia DR3 velocity scale is in satisfactory agreement with APOGEE when $G_0$ brighter than 14 mag, consistent with \citet{2022arXiv220605902K}. The LAMOST-released spectra are derived from the LAMOST 2D pipeline. The spectra are split into blue and red parts and are collected with two arms. The blue- and red-arm raw spectra are processed separately in the LAMOST 2D pipeline and are pieced together after flux calibration. During this progress the generation of errors is inevitable, especially for LAMOST low-S/N spectra.  More details of these systematic errors can be found in \citet{Luo2012} and \citet{2015MNRAS.448...90X}.

\subsection{Distance}
The distance parameter plays a vital role in the research of structure and evolution of the Milky Way. 
Li16 found that the color-magnitude relation for M giants based on stellar models may be biased, and devised a new empirical distance relation as follows:
\begin{equation}\label{dist_mo}
M_{J}=A_{1}[(J-K)_{0}^{A_{2}}-1]+A_{3}.
\end{equation}
Li16 also investigated metallicity and star formation history trends of this distance relation. The resulting distribution are significantly different in the Sgr and LMC systems due to their different star formation histories. So for the Milky Way star distances, we need refit this relation based on the star formation history of the Milky Way.

Considering that nearby stars observed by \gaia{} have more accurate distances and most of our stars are Milky Way disk stars, we cross-match our M giants with \gaia{} EDR3 distances \citep{2021AJ....161...147B} under the limitations for all stars closer than 4 kpc and parallax/parallax errors $>$ 5
(most are metal rich disk stars). Then we refitted equation \ref{dist_mo} and obtained the three new coefficients $A_{1}=6.07$, $A_{2}=-1.13$, and $A_{3}=-3.34$. This result is also well consistent with $M_{J,fit}$ relation for $[Fe/H]=0$ stars derived by \citet{2014ApJ...793...62S} with their stellar model.

\begin{figure*}
  \centering
  \includegraphics[angle=0,scale=0.45]{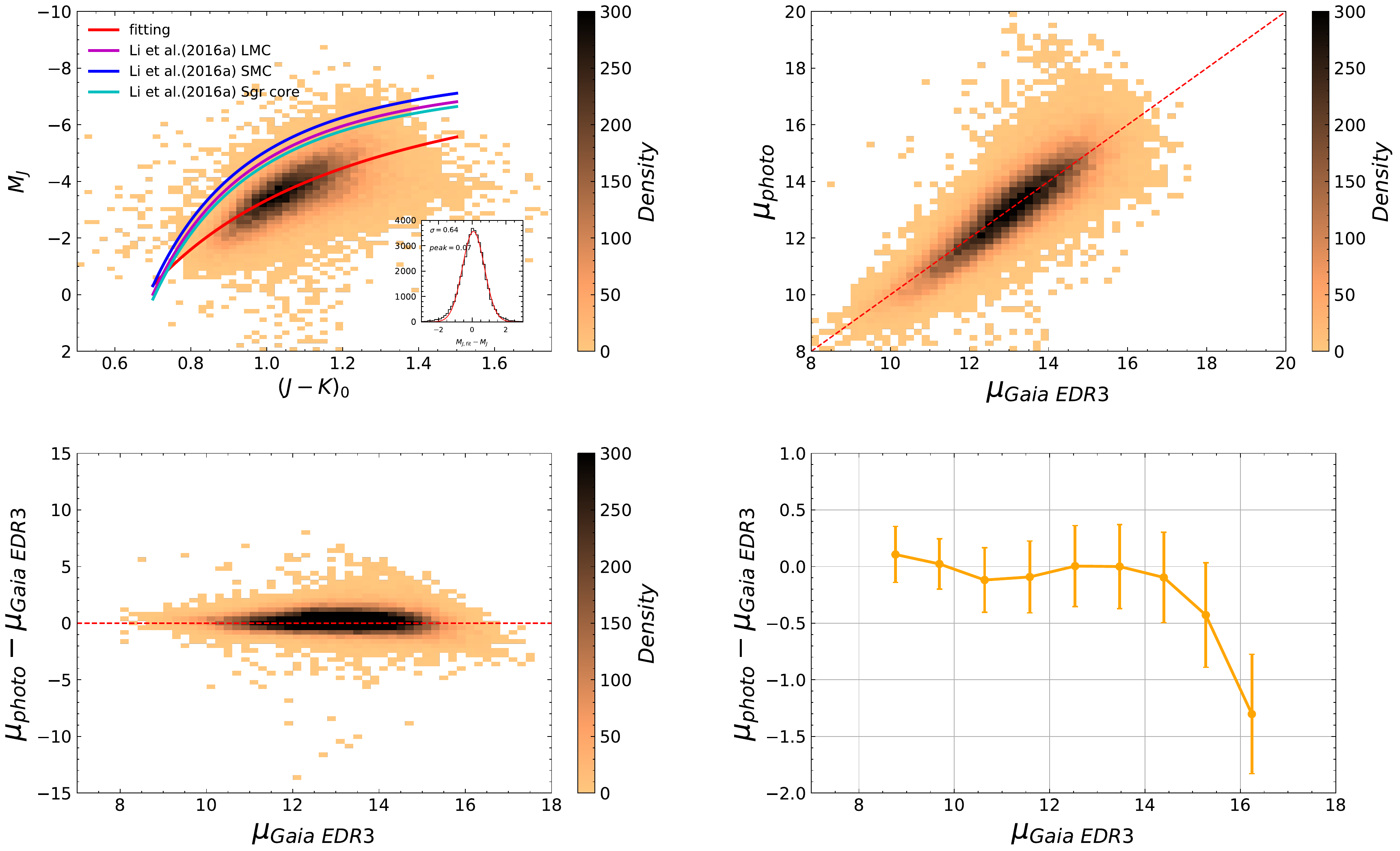}
  \caption{Top-left panel: $(J-K)_{0}$ vs. $M_J$ relation for the M-giants samples. The inset histogram shows the scatter around the $M_{J,fit}$ relationship, which has a dispersion of 0.64 dex. Top-right panel: distance modulus comparison between our new photometric derived relation vs. \gaia{} EDR3 distance derived by \citet{2021AJ....161...147B}. Bottom panel: further distance modulus comparison of the two different methods. The orange dots and error bars represent the median and standard deviation of each bin.}
  \label{dist}
\end{figure*}

In the top-left panel of Figure~\ref{dist}, the density map shows the selected M-giants distribution in a $(J-K)_{0}$ versus $M_J$ map. We compare our refitted $(J-K)_{0}$ versus $M_J$ relation (red line) to the Li16 relations (blue, purple and green lines). There are obvious discrepancies between Li16's and our relations. The inset histogram shows the scatter around the $M_{J,fit}$ relation. The color-absolute magnitude relation derived by Li16 uses the distances of the LMC, SMC, and the Sgr core region as the standard to estimate $M_J$ (e.g. all stars in the LMC region are assigned a distance of 51 kpc), while our $M_J$ is calculated using the more accurate distances for nearby stars derived from Gaia parallaxes; hence, our fitting shows more scatter. And also, we can see that the metallicities and star formation histories between the Milky Way and LMC, SMC, and Sgr systems are very different. 

Then we show the distance modulus distribution derived by our new relation versus the \gaia{} EDR3-calculated results in the top-right panel. As we can see, for these selected metal rich disk stars, these two methods derive the distance moduli consistently. In the bottom panels we show a comparison of these two distance modulus samples. It is easy to see that, when $\mu_{Gaia}$ larger than 15, the difference between $\mu_{photo}$ and $\mu_{Gaia}$ is steeply increasing.

So for our all M-giants sample, we separate it to two parts: the disk-like stars ($\mid$V - V$_{LSR} \mid <$ 180 km s$^{-1}$), for which we calculate the distances with our new fitted color-absolute magnitude relation, and halo-like stars ($\mid$V - V$_{LSR} \mid  >$ 180 km s$^{-1}$, where the most distant halo M-giant stars in our sample are Sgr stream members) for which we use the color-absolute magnitude relation derived from the Sgr core by Li16.

\subsection{Catalog Description}\label{catalog}

The complete catalog of M giants is provided in the electronic version of this paper, including 58,076 M giants. As described in \citet{2015AJ....150...42Z}, the classification results contain M giants with seven temperature subtypes from M0 to M6, as well as M dwarfs with 18 temperature subtypes from K7.0 to M8.5 and 12 metallicity subclasses from dMr to usdMp. The radial velocities and radial velocity errors of the M giants were calculated with the $laspec$ algorithm \citep{2021ApJS..256...14Z}. The spectral indices of CaH2, CaH3, and TiO5 were provided for measuring molecular features, which were defined by \citet{1995AJ....110.1838R} and \citet{2007ApJ...669.1235L}. In addition, our catalog includes photometry in optical bands $G$mag, $G_{bp}$, and $G_{rp}$ from \gaia{} EDR3, NIR bands $J$, $H$, and $K_S$ from 2MASS, the IR bands W1 and W2 from WISE, and distances as calculated by the M-giant color-absolute-magnitude relation. In Section~\ref{mstar}, the dereddened magnitudes ($Gmag_0$, $G_{bp0}$, $G_{rp0}$, $J_0$, $H_0$, $Ks_0$, $W1_0$, and $W2_0$) and absolute magnitude $M_G$ for each star were calculated, and all of these derived magnitudes are provided in our catalog. In Table~\ref{ctlg} we provide a description of each column of our catalog.

There are non-M-giant stars in the M-giants catalog. In order to mark those non-M-giant stars and further purify the M-giant-stars sample, we have added a column called ``SFLAG'' in the M-giants catalog, including ``gm'' as confirmed M giants, ``dm'' as M dwarfs, ``n'' as non-M-type stars, ``c'' as carbon stars, ``w'' as white dwarf binaries, and ``u'' as unconfirmed M-giant stars because of the lack of \gaia{} data. Table~\ref{sub} shows the number of sources in each labeled subsample.

\begin{table*}
\caption{Description of the M-giant catalog.}
\tabletypesize{\footnotesize}
\label{ctlg}
 \centering
\begin{tabular}{llcl}
\hline
Column &  Format  & Unit   & Description \\
\hline
filename & string & - & object's name from LAMOST DR9 \\
planid & string & - & object's planid from LAMOST DR9 \\
obsid & integer & - & object's obsid from LAMOST DR9 \\
R.A. & float & deg & object's R.A. in LAMOST DR9 (J2000)\\
decl. & float & deg & object's decl. in LAMOST DR9 (J2000)\\
CaH2 & float & -  & spectral index of the LAMOST spectrum \\
CaH3 & float & -  & spectral index of the LAMOST spectrum \\
TiO5 & float & -  & spectral index of the LAMOST spectrum \\
laspec\_rv & float & km s$^{-1}$ & radial velocity measured by the $laspec$ algorithm \\
laspec\_rv\_err & float & km s$^{-1}$ & radial velocity error measured by the $laspec$ algorithm \\
sn & float & - & mean S/N in the LAMOST spectrum \\
spty & string & -  & spectral subtype classified by the template-fitting algorithm \\
ebv & float & mag &  reddening from the 3D dust map\\
$M_G$ &  float & mag & $G$-band absolute magnitude from \gaia{} EDR3\\
$Gmag_0$ &  float & mag & dereddened $G$-band magnitude from \gaia{} EDR3\\
$G_{bp}$$_0$ &  float & mag & dereddened  $BP$-band magnitude from \gaia{} EDR3\\
$G_{rp}$$_0$ &  float & mag & dereddened  $RP$-band magnitude from \gaia{} EDR3\\
$J_0$ &  float & mag & dereddened $J$-band magnitude from 2MASS\\
$H_0$ &  float & mag & dereddened $H$-band magnitude from 2MASS\\
$K_{S0}$ &  float & mag & dereddened $K_S$-band magnitude from 2MASS\\
W1$_0$ &  float & mag & dereddened W1-band magnitude from WISE\\
W2$_0$ &  float & mag & dereddened W2-band magnitude from WISE\\
Dist$_{photo}$ & float & kpc &  distance calculated by the
M-giant color-absolute-magnitude relation \\
$[$M/H$]$ & float &  - & estimated photometric metallicity of M giants\\
$X$ & float & kpc & Galactocentric coordinate points to the direction opposite to that of the Sun \\
$Y$ & float & kpc & Galactocentric coordinate points to the direction of Galactic rotation \\
$Z$ & float & kpc & Galactocentric coordinate points toward the north Galactic pole   \\
$U$ & float & km s$^{-1}$ & Galactic space velocity along the $X$-axis   \\
$V$ & float & km s$^{-1}$ & Galactic space velocity along the $Y$-axis    \\
$W$ & float & km s$^{-1}$ & Galactic space velocity along the $Z$-axis   \\
SFLAG & string  & - & label of subsamples  \\
\hline
\end{tabular}
\begin{flushleft}
(This table is available in its entirely in FITS format.)
\end{flushleft}
\end{table*}

\begin{table*}
\caption{Number of sources of the labeled subsample in our M-type catalog.}
\label{sub}
 \centering
\begin{tabular}{lcrr}
\hline
 Source Type & ~Label &~~ M-giants Catalog \\
\hline
Confirmed gM-type stars   &   gm  &  44,036\\
Confirmed dM-type stars   &   dm  &  857\\
Non-M-type stars  & n & 1599\\
Carbon stars & c & 372\\
White dwarf binaries & w & 15\\
Unconfirmed M-type stars &  u &  11,197\\
\hline
Total & & 58,076\\
\hline
\end{tabular}
\end{table*}

\section{Structures Traced by M Giants}
\begin{figure*}
   \centering
   \includegraphics[angle=0,scale=0.33]{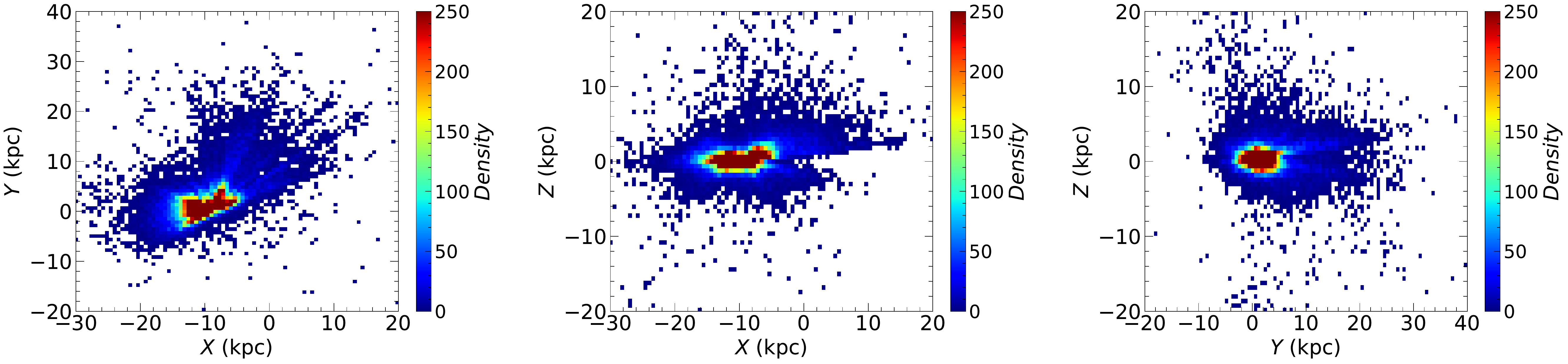}
   \caption{Galactocentric spatial distribution of the confirmed M giants in LAMOST DR9.}
   \label{xyz}
\end{figure*}

We calculated the Galactocentric coordinates ($X$, $Y$, $Z$) and space velocities ($U$, $V$, $W$) for all confirmed M giants. The solar position and the circular velocity at the solar location are adopted as $R_0$ = -8.34 kpc and $V_c$ = 240 km~s$^{-1}$, respectively \citep{2014ApJ...783..130R}. To correct for solar motion, we adopt the peculiar velocity of the Sun in the local standard of rest ($U_\odot$, $V_\odot$, $W_\odot$) = (11.1, 12.24, 7.25) km s$^{-1}$ \citep{2010MNRAS.403.1829S}. The spatial distribution of all our confirmed M giants showed in Figure~\ref{xyz}. 

\subsection{Sgr Stream}
The Sgr stream is the most prominent and extensive coherent stellar tidal stream in the Milky Way. A wide variety of stellar types have been used to trace Sgr tidal debris, such as main-sequence turn-off stars, blue horizontal branch (BHB) stars, M giants and so on \citep{2003ApJ...599.1082M,2019ApJ...874..138L,2019ApJ...886..154Y}. 

\begin{figure*}
 \centering  
   \includegraphics[angle=0,scale=0.75]{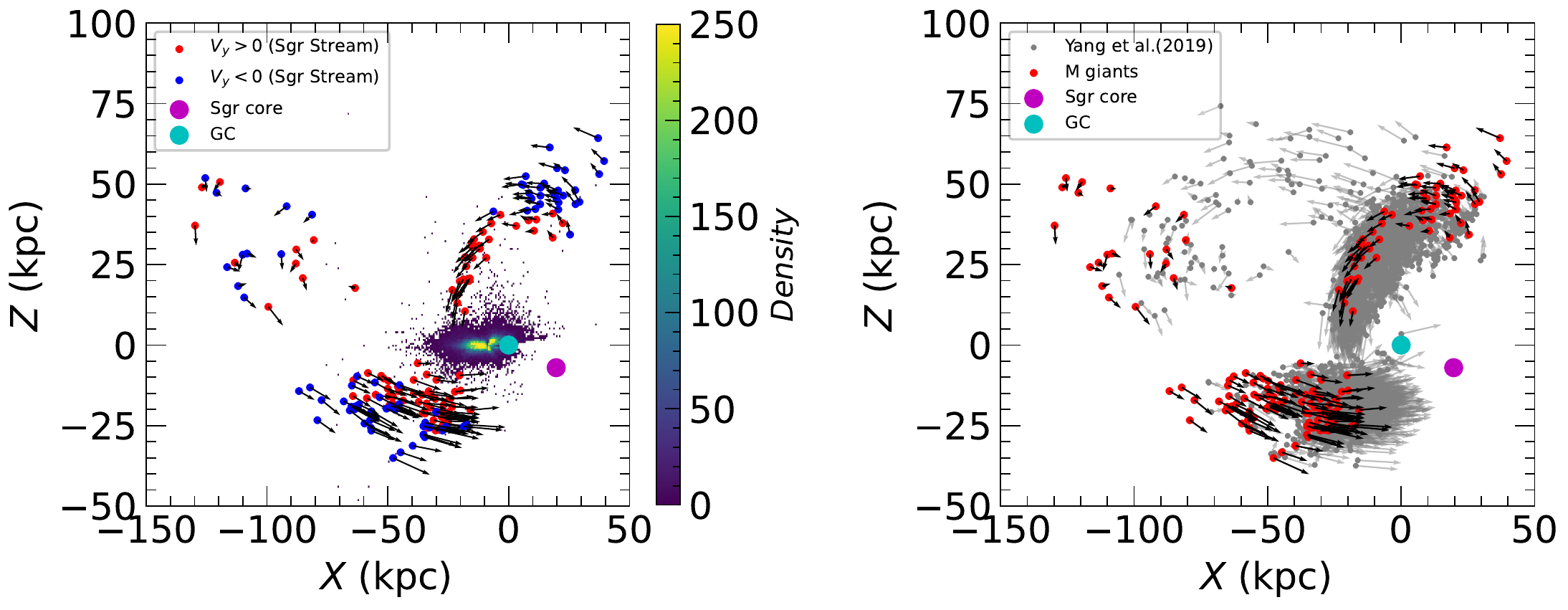}
   \caption{Spatial distribution of our Sgr stars. The arrows show the 3D velocity distribution for Sgr stars in the $X-Z$ plane (right handed). The purple dot marks the location of the Sgr dwarf galaxy (Sgr core). The Galactic center is marked with "GC" and a cyan dot. In the right panel, the gray dots represent Sgr stars from \citet{2019ApJ...886..154Y} and the red dots are our data.}
   \label{sgr}
\end{figure*}

To select candidate members of the Sgr stream from our updated M giants, we use the friends-of-friends (FoF) method in integrals-of-motion (IoM) space. The five IoM parameters are eccentricity $e$, semimajor axis $a$, the direction of the orbital pole ($l_{orb}$, $b_{orb}$), and the angle between the apocenter and the projection of the X-axis on the orbital plane $l_{apo}$, which is obtained through 6D information ($\alpha$, $\delta$, $d$, $hrv$, $\mu_{\alpha}$, $\mu_{\delta}$). Then we calculated the "distance" between any two stars in the normalized space of ($\alpha$, $\delta$, $d$, $hrv$, $\mu_{\alpha}$, $\mu_{\delta}$), and then finally used FoF to find out group stars that have similar orbits according to the size of "distance". The same method has been used in our previous works in \citet{2019ApJ...886..154Y} and \citet{2021ApJ...910...46L}. We finally obtained 183 Sgr member stars. There are 20 more Sgr stars than in the previous work of \citet{2019ApJ...874..138L} with LAMOST DR5 M giants. In the left panel of figure~\ref{sgr}, we show a density map of all identified M giants in the $X-Z$ plane, where the arrows show the 3D velocity distribution for the Sgr stars. In the right panel of Figure~\ref{sgr}, we compared our Sgr stars with the K giants in the previous study of \citet{2019ApJ...886..154Y}. As we can see, our sky distribution and 3D velocity distribution are well consistent with Sgr stars detected via K giants.

In this work, we only select candidate Sgr stars, then show their sky positions and orbits. In our following work, we will study the chemical abundance with $[$M/H$]$ and $[\alpha$/M$]$. Qiu et al.(2023, in preparation) predict the stellar parameters of M giants with the data-driven Stellar LAbel Machine model (SLAM) developed by \citet{2020ApJS..246....9Z}. 

\subsection{Disk Structures}

\begin{figure*}
     \includegraphics[angle=0,scale=0.32]{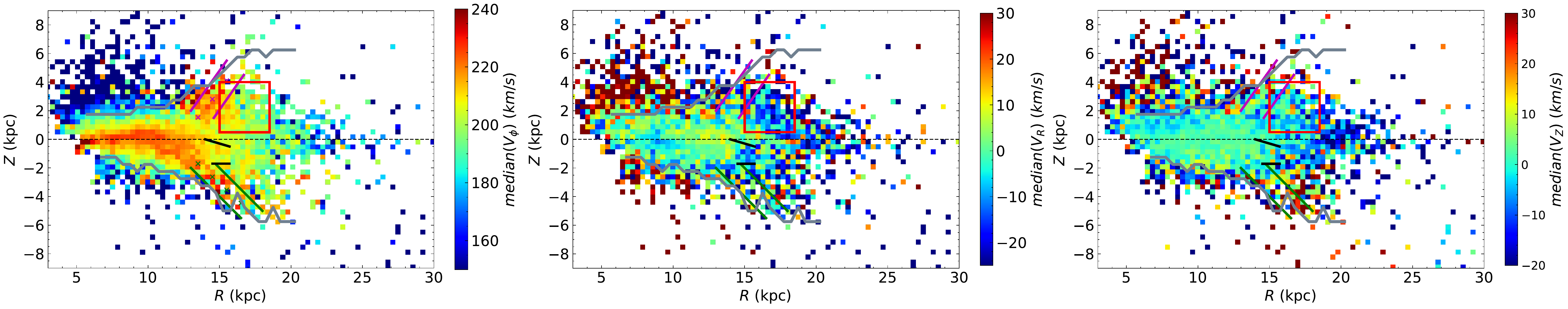}
  \caption{The median of $V_{\phi}$, $V_{R}$, and $V_{Z}$ for LAMOST M giants, as a function of ($R$, $Z$). The bounds of the "main branch", "north branch", and "south branch" are labeled by black lines, pink lines, and green lines, respectively. The north branch lies along ($R$, $Z$) = (13.5, 1.53), (14.5, 2.81), and (15.5, 4.1) kpc, and is labeled in the left panel by pink crosses. The south branch lies along ($R$, $Z$) = (13.5, -1.7), (14.5, -2.625), (15.5, -3.54), and (16.5, -4.46) kpc, and is labeled in the left panel by green crosses. The Monoceros area is labeled by a red square. The dark curves label the boundary of the region with median $V_{\phi}>160$ km s$^{-1}$ shown on the left.}
  \label{rz_velocity}
\end{figure*}

To check for substructures in the Galactic outer disk, we select disk stars with $\mid$V - V$_{LSR} \mid <$ 180 km s$^{-1}$ \citep{2004AJ....128.1177V, 2009MNRAS.399.1145S}. Then we construct heat maps of the median cylindrical velocities as a function of ($R$, $Z$), as shown in Figure~\ref{rz_velocity}. The bin size in our maps is ($\Delta R$, $\Delta Z$) = (0.35, 0.35) kpc.

Figure~\ref{rz_velocity} shows the median $V_{\phi}$, $V_{R}$, and $V_{Z}$ in the range of 
$3<R<30$ kpc and $-9<Z<9$ kpc. The left panel shows the median$V_{\Phi}$ for our selected M giants as a function of ($R$, $Z$). The disk and halo populations can be clearly distinguished, where the disk stars have high $V_{\Phi}$ values. \citet{2020ApJ...905....6X} found that beyond $R=11$ kpc, the most significant structure in disk is that the high $V_{\Phi}$ stars are split into three branches, namely the main, north, and south branches, found via K giants. And these important structures are also present in our plot found via M giants. In Figure~\ref{rz_velocity}, we can see the main branch clearly, which is found along the line from  ($R$, $Z$) = (12, −0.5) kpc to ($R$, $Z$) = (15, −1) kpc. The north branch is found along the slope from ($R$, $Z$) = (13, 1.5) kpc to (17, 5) kpc. The south branch is found along the line from ($R$, $Z$) = (11, −1) kpc to ($R$, $Z$) = (16, −3) kpc. There is an area with quite small median $V_{R}$ in the range of $15<R<18.5$ kpc and $0.5<Z<4$ kpc in the middle panel of Figure~\ref{rz_velocity}. 
In our work, the disk constructed by M giants cuts off around ($R$, $Z$) = (25, 0) kpc, 5 kpc further out than in the work of \citet{2020ApJ...905....6X} based on K giants. And we see that out to 25 kpc, the velocity of the disk is still asymmetric, as shown in the right panel of figure~\ref{rz_velocity}, where the median values of $V_{Z}$ and $V_{R}$ are negative when $R$ is larger than 18 kpc.
In our next work, we will study the detail asymmetry structures of the disk with our selected M dwarfs and M giants. 

\section{Summary}
 We have used the template-fitting pipeline derived in our previous work to identify and classify M dwarfs and M giants from LAMOST DR9 low-resolution spectra. A total of 58,076 M giants and 764,676 M dwarfs are provided. Using a \gaia{} EDR3 $M_G$ versus $(G_{bp}-G_{rp})_0$ diagram clearly shows that early-type stars, M dwarfs, and white dwarf binaries are misidentified in the M-giants sample. The contamination rate in the M-giants sample is 4.2\%.  In the published M-giants catalog, "SFLAG" is used to mark the confirmed M giants as well as other contamination sources. In particular, we found that the CaH spectral indices are efficient selection criteria for identifying carbon stars. Finally, about 372 carbon stars were selected from the M-giants sample, and further confirmed through visual inspection of the LAMOST spectra. 

We measured the radial velocities of our M giants with the cross-correlation-based $laspec$ algorithm derived by \citet{2021ApJS..256...14Z}. The velocity offset and scatter with Gaia DR3 and APOGEE DR17 are around $\sim$ 1 km s$^{-1}$ and $\sim$ 4.6 km s$^{-1}$, respectively. When $G_{0}$ is fainter than 14 mag, the discrepancy is larger than $\sim$ 4 km s$^{-1}$. 
 
We refit the color$-$metallicity relation with updated APOGEE metallicities. Furthermore, we also refit the color-magnitude relation using the distances of nearby stars measured bu \gaia{} \citep{2021AJ....161...147B}, whose metallicities and star formation histories are more like disk stars. So for our M giants,
for the disk-like stars ($\mid$V - V$_{LSR} \mid <$ 180 km s$^{-1}$), we calculate the distances with our new fitted color-absolute magnitude relation, and for the halo-like stars ($\mid$V - V$_{LSR} \mid  >$ 180 km s$^{-1}$), we use the color-absolute magnitude relation derived using Sgr core stars by Li16 to calculate the distances.

Using the FoF method in IoM space, we select 183 Sgr stars, and trace the Sgr stream in 6D phase space across the sky coverage. As figure~\ref{sgr} shows, our M-giant Sgr stars are well consistent with the K giants studied in a previous study, both in terms of sky position and orbit motion.

We also traced the substructures in the Galactic outer disk. As figure~\ref{rz_velocity} showing, our M giants also clearly represent the substructures "main branch", "north branch", and "south branch" found in \citet{2020ApJ...905....6X}. It is worth noting that in our work, the disk constructed by M giants cuts off around ($R$, $Z$) = (25, 0) kpc, which is 5 kpc further than in the work of \citet{2020ApJ...905....6X} based on K giants. And our M giants show that out to 25 kpc, the velocity of the disk is still asymmetric, because the median $V_{Z}$ and $V_{R}$ values are negative when R larger than 18 kpc. 



\acknowledgments

We would like to acknowledge National Key RD Program of China No. 2019YFA0405504, the NSFC under grants 12273027, the Sichuan Youth Science and Technology Innovation Research Team (grant No. 21CXTD0038)，
and the Innovation Team Funds of China West Normal (grant No. KCXTD2022-6).
J.Z. would like to acknowledge the NSFC under grants 12073060, and the Youth Innovation Promotion Association CAS. J.-R.S. would like to acknowledge the National Natural Science Foundation of China under grant Nos. 12090040, 12090044, and 11833006. The Guoshoujing Telescope (the Large Sky Area Multi-Object Fiber Spectroscopic Telescope LAMOST) is a National Major Scientific Project built by the Chinese Academy of Sciences. Funding for the project has been provided by the National Development and Reform Commission. LAMOST is operated and managed by the National Astronomical Observatories, Chinese Academy of Sciences.

This work has made use of data from the European Space Agency (ESA) mission Gaia (\url{https://www.cosmos.esa.int/gaia}), processed by the Gaia Data Processing and Analysis Consortium (DPAC, \url{https://www.cosmos.esa.int/web/gaia/dpac/consortium}). Funding for the DPAC has been provided by national institutions, in particular the institutions participating in the Gaia Multilateral Agreement. Our work also made use of data products from the
Wide-field Infrared Survey Explorer \citep[ALLWISE Source Catalog;][]{2013wise.rept....1C}, which is a joint
project of the University of California, Los Angeles,
and the Jet Propulsion Laboratory/California Institute
of Technology, funded by the National Aeronautics and
Space Administration.



\end{CJK*}

\end{document}